\documentclass[11pt,twoside]{article}  
\usepackage{apn3conf}

\def\nii{[N {\sc ii}]}
\def\sii{[S {\sc ii}]}
\def\oii{[O {\sc ii}]}
\def\oiii{[O {\sc iii}]}
\def\ha{H$\alpha$}
\begin{document}   

\title{Characterizing Low-Ionization Structures in PNe}
\titlemark{Low-Ionization Structures---LIS}

\author{Denise R. Gon\c calves}
\affil{Instituto de Astrof\'{\i}sica de Canarias, 
       \\E-38205 La Laguna, Tenerife, Spain, Email: denise@ll.iac.es}

\contact{Denise R. Gon\c calves}
\email{denise@ll.iac.es}

\paindex{Gon\c calves, D. R.}

\authormark{D. R. Gon\c calves}


\begin{abstract}          
Fifty five planetary nebulae containing micro-struc\-tures of 
low-ionization (LIS) are analyzed in this review in terms of LIS morphology, 
kinematics, and physical and excitation properties. We attempt to address 
the issue of their origin through the comparison of LIS properties 
with the main shells of the nebulae, as well as the contrast with 
the theoretical model predictions. We finally conclude that, while 
LIS morphology and kinematics can be reasonably accounted for by the 
available theoretical models, they cannot explain the LIS 
physical/excitation properties, unless we agree that evolved PNe are 
expected to show neither shock-excited LIS nor significant density 
contrasts relatively to their environments. Some evidence for the latter 
ideas is actually presented in this paper.
\end{abstract}

\section{Introduction}

Low-ionization `microstructures' of planetary nebulae (PNe) are those 
structures that are much more prominent in \nii, \sii, and \oii\ than 
in the \oiii\ or \ha\ emission lines. The low-ionization structures  
(LIS, Gon\c calves et al.\ 2001) appear with different morphological 
and kinematical properties in the form of 
pairs of knots, filaments, jets, or isolated features moving with 
supersonic velocities through the large-scale components in which 
they are located, or instead as structures with the above morphologies 
but with low velocities that do not differ substantially from that of 
the main nebula\footnote{The fast, low-ionization emission regions, 
FLIERs (Balick et al.\ 1993) and bipolar, rotating, episodic jets, 
BRETs (L\'opez et al.\ 1995) are particular 
types of LIS, i.e., high-velocity knots and jets.}. For the sake of 
conciseness, the complete table listing all known LIS will not be
reproduced here; however, readers are referred to the web page at  
www.iac.es/galeria/denise/PNe\_LIS.html where the morphology and
kinematics of LIS, as compared to those of the main PN bodies, are shown. 
This table, adapted from Gon\c calves et al. (2001), also includes those 
`new' objects whose LIS were identified after 2001. 

It is important to keep in mind the number of PNe that are known to possess 
LIS. These amount to 55 objects. If we consider the number of 
Galactic PNe imaged with at least one filter of higher and another of 
lower ionization (Balick 1987; Schwarz et al. 1992; Manchado et al.\ 
1996), we end up with 527 PNe. Therefore, about 10\% of the Galactic PNe are 
actually confirmed to have LIS.

\section{Types of LIS from their Morphology and Kinematics}

\setcounter{footnote}{3}

\begin{figure}
\epsscale{.50}
\plotone{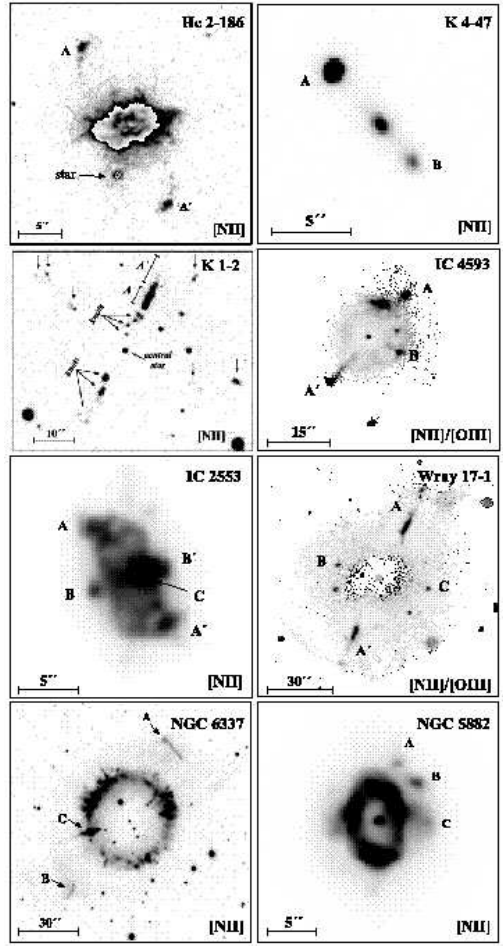}
\caption{Examples of the different types of LIS. From top to bottom, we 
show two PNe with low-ionization pairs of jets, jetlike pairs, pairs of 
knots, and finally the isolated structures.} \label{fig1-review}
\end{figure}

Many authors, as can be noted from the above-cited table, have been working 
on the 
characterization of LIS. So the reader is referred to the list in the 
table in order to obtain information on the morphology/kinematics for 
each of the objects listed. On the other hand,  Gon\c calves 
et al. (2001) offers a detailed study concerning the morphological and
kinematic classification of LIS as compared to theoretical model 
predictions, where the authors gather together 50 PNe containing LIS, with 
all their 
different types (see Figure~\ref{fig1-review}. Also check Tables 2 to 5 of 
Gon\c calves 
et al.\ (2001) for the kinematical ages, positions and orientations of 
LIS with respect to the rim and the major axes of the PN main bodies). 
The most relevant conclusions, reached from the morphology and kinematics 
of LIS, are shown below.

\subsection{Pairs of Jets}
Most of the properties of jets in NGC~7009, NGC~6891, and NGC~3918 can be 
explained by HD and MHD interacting stellar wind (ISW) models. The observed 
linear v$_{\rm exp}$ $\times$ d (distance from the center) would favor the MHD 
(Garc\'\i a-Segura et al.\ 1999) and the stagnation jet (Steffen et al. 
2001) models. K~4-47, M~1-16 and Fg~1 would better be explained by 
accretion disk jet models (Blackman et al. 2001); while Hb~4, NGC~6210, 
and NGC~6543, showing jets which are kinematicaly younger than the main 
PNe shells, do not fit within the predictions of any model. Some jets 
have large (say 90$^{\circ}$) jet--nebula angles (Hb~4, NGC~6210, and 
NGC~6884). It is still not clear if the magnetic/rotation axis and 
polar axis misalignments (Garc\'\i a-Segura \& L\'opez 2000; 
Blackman et al. 2001) can account for large tilt angles like these.
                           
\subsection{Jetlike Pairs}
IC~4593, He~2-429, NGC~6881, K~1-2, Wray~17-1, and probably NGC~6751 
show features that are very much like jets, but that are moving with 
velocities which do not differ substantially from those of the PN main 
bodies (see Corradi et al.\ 1997, 1999; Guerrero et al.\ 1999). Models 
only predict high-velocity jets instead of jetlike features, with the 
exception of the {\it transient} low-velocity collimated micro-structures 
of the Soker (1992), R\'o$\dot{\rm z}$yczka \& Franco (1996), and 
Steffen et al.\ 
(2001) models.

\subsection{Pairs of Knots}
Symmetrical high-velocity pairs of knots can, in principle, be accounted for 
by HD or MHD interacting stellar winds, accretion disk winds, or 
stagnation zone models, since models for the jet formation can also easily 
explain the origin of pairs of knots. The non-detection 
of the outward-facing bow shocks that should be associated with these 
features might be caused by instabilities (HD/radiative; Soker \& Reveg 
1998). On the other hand, pairs of low-velocity knots are less studied 
theoretically. We suggest that fossil AGB knots, implying a very peculiar AGB 
mass-loss geometry, 
could be responsible for at least some of the low-velocity pairs of knots. 

\subsection{Isolated LIS}
Isolated LIS may be formed by in situ instabilities and/or by  
fossil AGB mass-loss inhomogeneities. However, from their 
positions with respect to the rim (the shell formed by the interaction 
between the AGB and the post-AGB winds), they are not related to 
dynamical instabilities because of the action of the fast post-AGB wind 
(see Table~5 of Gon\c calves et al.\ 2001). The rocket effect (Mellema et al.\
 1998) 
could, in some cases, explain the peculiar velocities of the high-velocity
isolated LIS.

\section{The Physical Parameters and the Excitation Properties of LIS}
From the above, it seems clear that at least part of the LIS 
characteristics can be reasonably explained by the different models 
intended to account for their formation, at least in terms of morphology 
and kinematics. Now, which are the observables for further comparison 
with models? How we can go further in distinguishing between the various 
physical processes for the origin of 
LIS?

\begin{figure}
\epsscale{.80}
\plotone{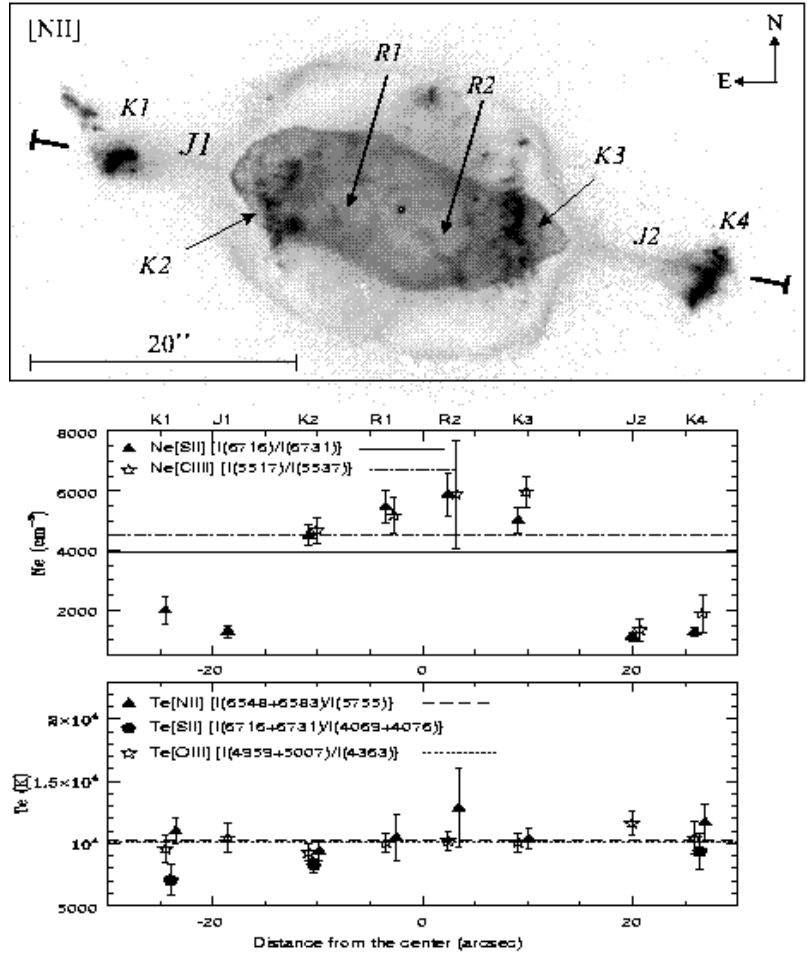}
\caption{The HST archive \nii\ image of NGC~7009 with its many structures, 
and the $T_{\rm e}$ and $N_{\rm e}$ associated with these structures.} 
\label{fig2-review}
\end{figure}

Densities or density contrasts of LIS are reported in the literature 
for many PNe individually. Balick et al.\ (1994) and Hajian et al.\ (1997) 
dealt with physical properties in a group of PNe with 
high-velocity pairs of knots (FLIERs). We are now trying to answer issues 
such as: i)~whether the density contrast plays a 
role on the low-velocity LIS; ii)~whether the low-density pairs could be 
features slowed down by the medium (since they are more easily stopped); and 
iii)~whether the high-density knots might originate in mass-loss episodes 
prior to PN formation. We are dealing with these questions by analyzing 
not only FLIERs but also the other types of LIS, i.e., the low-velocity ones.

What about the excitation mechanisms of the low-ionization structures?
Should we observe shocked--excited emission from the high-velocity LIS 
(Dwarkadas \& Balick 1998; Soker \& Reveg 1998)? 

Let us analyze the case of NGC~7009 (which shows one of the best known  
jets in PNe) and then compare its characteristics with those of other 
PNe with low- and high-velocity pair of knots, as well as jets.

\subsection{$T_{\rm e}$, $N_{\rm e}$ and the Excitation of NGC~7009}

The physical, chemical, and excitation properties of NGC~7009, including its 
pair of jets (J1, J2), inner (K2, K3) and outer (K1, K4) pairs of knots, and 
the rim (R1, R2), were recently studied by Gon\c calves et al.\ (2003); see 
Figure~\ref{fig2-review}. This is one of the PNe in our sample observed with 
the 2.5 m~INT (ORM, La Palma, Spain), in August 2001. Our spectra were taken 
at P.A. = 79$^{\circ}$ (slit width = 1$''$.5, slit length = 4$'$) and 
cover the optical range from 3700~\AA\ to 6750~\AA, with 3.3~\AA~pix$^{-1}$
and 0$''$.70~pix$^{-1}$ resolutions (see also Gon\c calves et al., this 
volume).

As is clear from Fig.~\ref{fig2-review}, the $T_{\rm e}$ throughout the 
nebula is remarkably constant, $T_e$\oiii\ $\approx$ 10\,200~K; the bright 
inner rim and inner pair of knots have similar densities of 
$N_e$\sii\ $\approx 6000$~cm$^{-3}$, whereas a much lower density of  
$\sim$ 1500~cm$^{-3}$ is derived for the outer knots, as well as for the 
jets. This implies that the outer pair of knots, being as dense as the jets, 
cannot originate from the matter swept up by jets, as expected from some 
model predictions (see discussion below).

We use two relevant diagnostic diagrams (Phillips \& Cuesta 1999) to 
investigate the excitation of the selected features in 
NGC~7009 (Figure~\ref{fig3-review}). These diagrams separate the zones 
of radiatively excited emission lines (the PN zone) from the zone mainly 
excited by shocks. It is clear that all eight structures in NGC~7009 
are mainly radiatively excited by the central star of the PN.

\begin{figure}
\epsscale{0.50}
\plotone{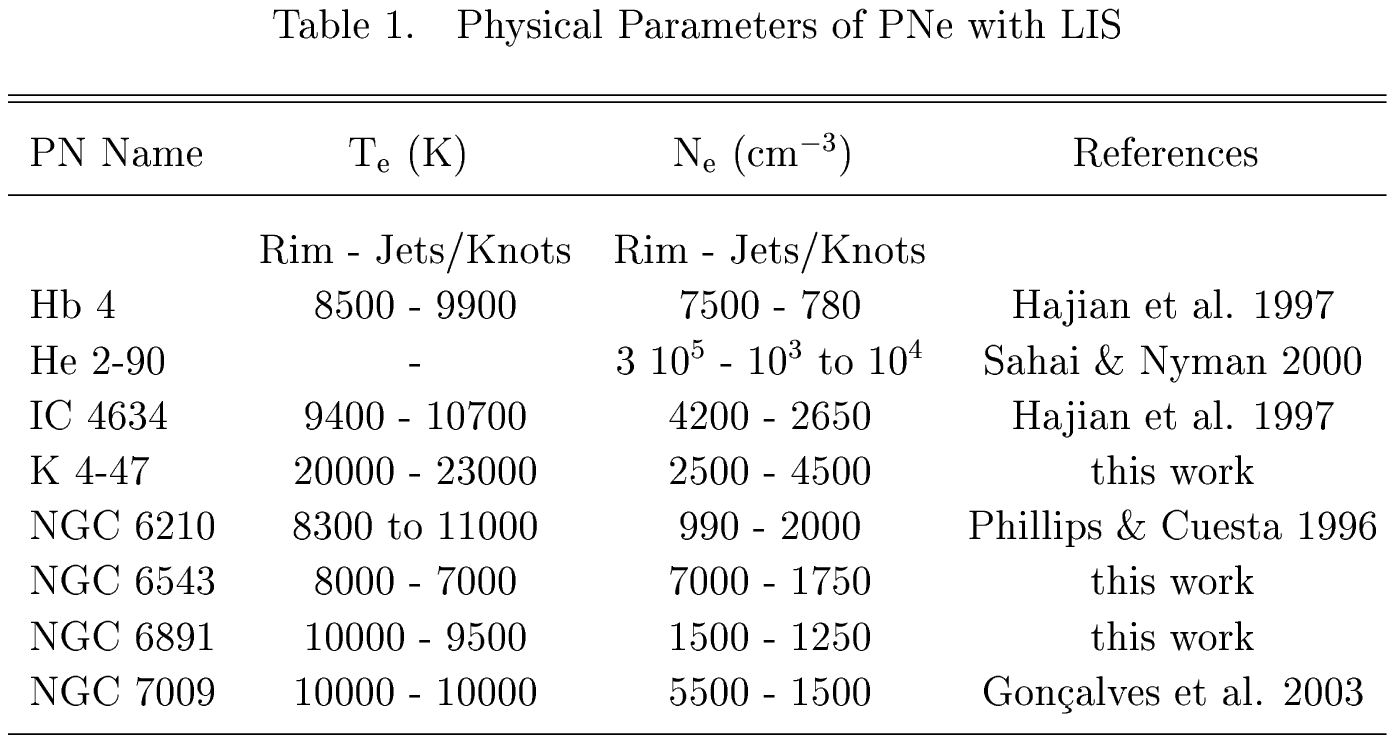}
\end{figure}

\subsection{$T_{\rm e}$, $N_{\rm e}$ and the Excitation of Other PNe with LIS}

In Table~1 we collect the $T_{\rm e}$ and $N_{\rm e}$ of some PNe with 
jets/knots. In the second and third columns, these properties for the 
main shell (rim) are shown, followed by those for the 
jets/knots in the same PNe. Note, from the table, that most jets/knots 
are between 2 and 10 times less dense than the nebular main bodies 
or their surroundings (the exceptions being K~4-47 and NGC~6210). 
What does this mean? Have the density contrasts 
been erased by the knots' expansion because of the ionization process, as 
suggested by Soker \& Reveg (1998)?

\begin{figure}
\epsscale{.90}
\plotone{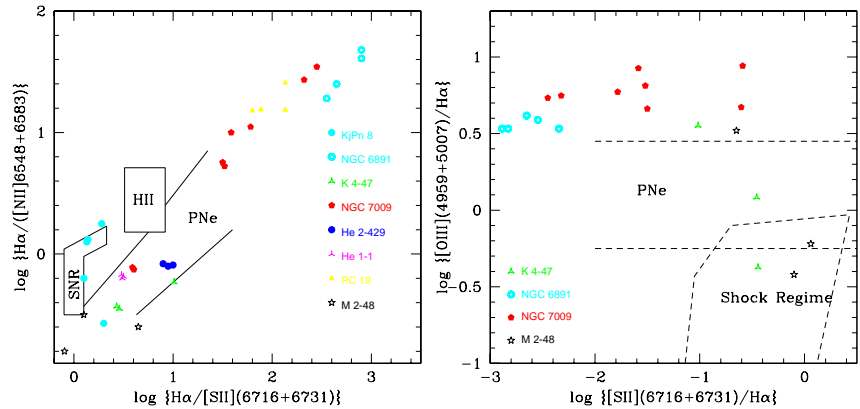}
\caption{Diagnostic diagrams showing the position of the emission-line 
ratio characteristics of shock-excited objects (SNR, shock regime) and 
the photoionized (PNe, HII) ones. References for the data: Fg~1, V\'azquez 
et al.\ (1998); PC~19, He~2-249, and He~1-1, Ben\'\i tez et al.\ (2002); 
Kj~Pn~8, 
L\'opez et al.\ (1995); NGC~7009, Gon\c calves et al.\ (2003); and NGC~6891 
and K~4-47, this work.} \label{fig3-review}
\end{figure}

In the diagnostic diagrams, presented in the previous section, we actually  
put together many PNe. Please take into account that the microstructures  
in Figure~\ref{fig3-review} are real jets, jetlike systems, and both high- 
and low-velocity pairs of knots. The first important result we can 
extract from these plots is that low-velocity LIS (in He~2-429 and He~1-1) 
cannot be separated from the high-velocity ones (all the others), most 
of them being distributed in the PN zone (an emission characteristic 
of photoionization instead of shock excitation) or above it. Second, 
only a few LIS (those of Kj~Pn~8, K~4-47, and M~2-48) are mainly excited 
by shocks. These three PNe appear to be the youngest in the sample. Are 
the jets/knots in the other, more evolved, PNe `relaxed' in the sense that 
the once shock excited LIS were reached by the central star ionization 
front, which came to be the most relevant excitation process? 

\section{Discussion and Conclusions}
In summary, what do we really know about LIS in PNe?
 From the observations: 

\begin{enumerate}
\item LIS appear as pairs of jets, knots, filaments, and 
      jetlike features, or as isolated systems; 
\item Sometimes LIS expand with the rim, shells or haloes in which    
      they are embedded, but sometimes they are much faster than the  
      main PNe components;
\item They are spread indistinctly within all the   
      morphological classes of PNe;
\item In general, they do not have an important density contrast with   
      respect to the main bodies; and 
\item Most LIS systems studied up to now are mainly photoionized.
\end{enumerate}
 
Concerning the comparison of LIS properties with the theoretical predictions, 
some of their characteristics appear hard to explain (see Balick \& Dwarkadas 
1998; Gon\c calves et al. 2001, 2003; Balick \& Frank 2002). However, the 
origin of part of these systems (from their morphology and kinematics) can 
be reasonably understood via ISW models, in single stars or binaries, 
with or without magnetic fields, precession, and wobbling. 

The fact that the expected shock excitation of jets and other high-velocity 
LIS is not usually observed could mean that jets/knots are relaxed 
systems in the sense that shock excitation and high density contrasts 
are no longer present because LIS were already reached 
by the energetic photons of the post-AGB central star, and/or affected by 
local instabilities (Dopita 1997; Miranda et al. 2000; Soker \& Reveg 1998). 
Finally, it is remarkable that both shock excited LIS and LIS with high 
density contrasts are found in young PNe.  
This may imply that jetlike systems and low-velocity pairs of knots 
are the remnant of real jets and high-velocity pairs of knots, 
which suffered the effects of drastic slowing down and photoionization, 
being therefore, found in more evolved PNe.

\end{document}